\documentclass[11pt]{article}
\usepackage{moriond,epsfig}

\def\lesssim{\mathrel{\mathpalette\vereq<}}
\def\gtrsim{\mathrel{\mathpalette\vereq>}}
\makeatletter
\def\vereq#1#2{\lower3pt\vbox{\baselineskip1.5pt \lineskip1.5pt
\ialign{$\m@th#1\hfill##\hfil$\crcr#2\crcr\sim\crcr}}}
\makeatother

\def\be{\begin{equation}}
\def\ee{\end{equation}}
\def\bea{\begin{eqnarray}}
\def\eea{\end{eqnarray}}

\begin{document}
\vspace*{4cm}
\title{DYNAMICAL (SUPER)SYMMETRY BREAKING}

\author{ HITOSHI MURAYAMA }

\address{Department of Physics, Center for Theoretical Physics\\
University of California, Berkeley, CA 94720, USA}

\maketitle\abstracts{Dynamical Symmetry Breaking (DSB) is a concept 
theorists rely on very often in the discussions of strong dynamics, 
model building, and hierarchy problems.  In this talk, I will discuss 
why this is such a permeating concept among theorists and how they are 
used in understanding physics.  I also briefly review recent progress 
in using dynamical symmetry breaking to construct models of 
supersymmetry breaking and fermion masses.}

\section{Hiearchy Problems, Stability of Hiearchy}

There are many hierarchy problems in the Standard Model.  A hierarchy 
is a puzzle if there is no apparent reason for one number to be much 
smaller than the other.  To understand them is one of the main 
applications of the DSB. Here is the list of most important hierarchy 
problems:
\begin{enumerate}
	\item  Smallness of the electroweak scale: $v / M_{\rm Planck} \sim 
	10^{-17} \ll 1$.

	\item  Fermion mass hierarcy: $m_{e}/m_{t} \sim 2\times 10^{-6} \ll 1$.

	\item Flatness of the Universe.  In order for the Universe now to 
	be as flat as observed, it must have been extremely flat in the 
	Early Universe at temperature $T$: $|\Omega-1| \lesssim 3 \times 
	10^{-16} (1~{\rm MeV}/T)^{2} \ll 1$.

	\item  Cosmological constant.  Using the current ``cosmic 
	concordance'' number, $\Lambda/M_{\rm Planck}^{4} \sim 10^{-123}$. 

	\item  Ultimate hierarchy: H.M. $\ll$ Bill Gates.
\end{enumerate}
None of the above hierarchies have obvious reasons for them, and our 
very existence crucial depend on them.

A very important question about a hiearchy is if it is stable.  
Certainly for the case of Bill Gates, it is; once there is an 
accumulated wealth, it reproduces itself and the hiearchy does not 
collapse easily (or even tends to grow).  If there were any reasonable 
explanation behind a hierarchy, the hierarchy should be stable; 
otherwise it is either transient, accidental, or initial condition 
dependent and there is no hope to understand it.  Only when the 
hiearchy is stable against small perturbations, we can get started to 
ask the question why.

Here is one examle of an unstable hierarchy: the electron mass in 
classical electromagnetism.\cite{INS}  Since an electron generates a Coulomb 
field around it, and it feels its own Coulomb field, there is a 
self-energy for the electron
\begin{equation}
	\Delta E \sim + \frac{e^{2}}{4\pi \epsilon_{0} r_{e}},
\end{equation}
where $r_{e}$ is the ``size'' of the electron.  Thanks to Einstein, 
mass is nothing but the rest energy, and the observed mass is the sum 
of the ``bare'' mass and the rest energy
\begin{equation}
	(m_{e} c^{2})_{\rm obs} = (m_{e} c^{2})_{\rm bare} + 
		\frac{e^{2}}{4\pi \epsilon_{0} r_{e}}.
\end{equation}
The problem is that the mass is linearly divergent in the pointlike 
limit $r_{e} \rightarrow 0$.  In fact, we know $r_{e} \lesssim 
10^{-17}$~cm experimentally, and the above equation would look 
numerically like
\begin{equation}
	0.511 = -9999.489 + 10000.000.
\end{equation}
If the ``bare'' mass was different by 1\%, the electron mass becomes 
200 times larger!  The smallness of the electron mass $m_{e} c^{2} \ll 
e^{2}/(4\pi \epsilon_{0} r_{e})$ is the hierarchy problem in this 
case, and this is an unstable hierarchy; a small perturbation destroys 
the hierarchy.  This implies that the classical electromagnetism is 
not applicable for distances below $e^{2}/(4\pi \epsilon_{0} m_{e} 
c^{2}) \sim 10^{-13}$~cm.

It turned out that the quantum mechanics and the discovery of 
anti-particles made the hierarchy stable.  The Coulomb self-energy 
discussed above can be depicted as a diagram where the electron emits 
a virtual photon (Coulomb field) and reabsorbs it (feels it).  This 
gives a positive self-energy $e^{2}/(4\pi \epsilon_{0} r_{e})$.  But 
now that we know the world is quantum mechanical and there exists the 
anti-particle of the electron, namely positron, we have to consider 
the following funny process.  The vacuum constantly fluctuates to 
produce a pair of electron-positron together with a photon, where 
these particle survive within the time allowed by the uncertainty 
principle $\Delta t \sim \hbar/\Delta E = \hbar/(2 m_{e}c^{2})$ and 
annihilate back to the vacuum.  When you put an electron in the 
vacuum, the electron sees the fluctuation, and sometimes it 
annihilates the positron in the vacuum fluctuation.  Then the electron 
that was originally in the quantum fluctuation remains as a real 
electron.  If you calculate the contribution of this process to the 
self-energy, you find $-e^{2}/(4\pi \epsilon_{0} r_{e})$.  The leading 
linearly divergent pieces exactly cancel between these two 
contributions and the self-energy ends up with a sub-leading piece:
\begin{equation}
	(m_{e} c^{2})_{\rm obs} = (m_{e} c^{2})_{\rm bare}
		+ \frac{3\alpha}{4\pi} (m_{e} c^{2})_{\rm bare}
		\log \frac{\hbar}{m_{e}c r_{e}}.
\end{equation}
Now that the self-energy depends only logarithmically on $r_{e}$, the 
correction to the electron mass is only 9\% even if the electron is as 
small as we can imagine: the Planck size $\sim 10^{-33}$~cm where the 
quantum field theory breaks down anyway.  Because the self-energy is 
proportional the bare mass itself, a 1\% perturbation in the bare mass 
results in a 1\% change in the observed mass, and the hierarchy is 
stable.  Once the hierarchy $(m_{e} c^{2})_{\rm bare} \ll e^{2}/(4\pi 
\epsilon_{0} r_{e})$ is set, it stays including the self-energy 
contribution.  The reason behind the stability is the chiral symmetry 
$\psi_{e} \rightarrow e^{i\theta \gamma_{5}}\psi_{e}$, which is exact 
in the massless electron limit, but is explicitly broken by the finite 
electron mass $m_{e}\neq 0$.  This is why the self-energy, which 
violates chiral symmetry, comes out proportional to the violation of 
chiral symmetry in the theory, the electron mass itself.  The 
hierarchy is made stable by doubling the number of particles.

We have learned that the smallness of the electron mass, or in general 
fermion mass hierarchy, is stable.  Therefore we can hope to understand 
the hierarchy.  Indeed one can write down models which naturally 
explain fermion mass hierarchy using approximate flavor symmetry as we 
will come back to later.

The problem which has been concerning many theorists is that the 
smallness of the electroweak scale is not a stable hierarchy.  In the 
Standard Model, we have the Higgs potential
\begin{equation}
	V = m_{H}^{2} |H|^{2} + \lambda |H|^{4}
\end{equation}
and the electroweak symmetry is broken spontaneously in the vacuum 
because the coefficient of the quadratic term is negative $m_{H}^{2} < 
0$.  The electroweak scale is given by $v=\langle H \rangle = 
\sqrt{-m_{H}^{2}/2\lambda} = 174$~GeV. A self-energy diagram for the Higgs 
boson produces a virtual pair of top-quark which annihilates back into 
the Higgs boson.  This diagram gives the correction
\begin{equation}
	\Delta m_{H}^{2} = -\frac{3h_{t}^{2}}{16\pi^{2}} \frac{1}{r_{H}^{2}},
\end{equation}
where $h_{t} \approx 1$ is the top quark Yukawa coupling and $r_{H}$ 
the ``size'' of the Higgs boson.  Because $\lambda \gtrsim 1$ violates 
perturbative unitarity, $|m_{H}^{2}| \lesssim (174~\mbox{GeV})^{2}$ for the 
Standard Model to be well-defined.  Then following the same logic as 
in the case of the electron mass, the Standard Model is not 
applicable to distance scales below $4\pi/\sqrt{3} v \sim 1$~TeV! If it is 
applied to much shorter distance scale, the smallness of the 
electroweak scale is spoiled by small perturbations.

The idea of supersymmetry is that the instability of the electroweak 
scale is solved by doubling the number of particles again.\cite{INS}
In supersymmetry, we introduce superpartners to every particle in the 
Standard Model.  The superpartner of the top quark $\tilde{t}$ 
contributes also to the self-energy of the Higgs boson.  If you 
calculate it, you find
\begin{equation}
	\Delta m_{H}^{2} = +\frac{3h_{t}^{2}}{16\pi^{2}} \frac{1}{r_{H}^{2}},
\end{equation}
and the leading quadratically divergent pieces cancel.  The total 
self-energy correction is then given by
\begin{equation}
	\Delta m_{H}^{2} = - 6 \frac{h_{t}^{2}}{16\pi^{2}}
		(m_{\tilde{t}}^{2}-m_{t}^{2}) \log \frac{\hbar}{m_{\tilde{t}}cr_{H}}.
\end{equation}
Hiearchy is now stable even for truly elementary (Planck-sized) Higgs 
boson, {\it as long as}\/ $m_{\tilde{t}} \lesssim \mbox{a few 100 
GeV}$.  This is why supersymmetry at sub-TeV scale is interesting; we 
can get {\it started}\/ to ask why the electroweak scale is so small.  
In the Minimal Supersymmetric Standard Model, $m_{H}^{2}$ is 
positive in the supersymmetric limit, and the electroweak symmetry cannot 
be broken; only after supersymmetry is broken, $m_{H}^{2}$ can be 
negative.  Therefore, breaking of supersymmetry induces the 
electroweak symmetry breaking.  The question we would like to ask then
is why the supersymmetry breaking scale is so low.

\section{Dynamical Symmetry Breaking at Work}

We have discussed that both the fermion mass hierarchy (thanks to 
chiral symmetry) and the smallness of the electroweak scale (if 
supplemented by supersymmetry) are stable, so that we are now entitled 
to ask why these hiearchies exist in nature.  But neither chiral 
symmetry nor supersymmetry explains the origin of hierarchy.  Both of 
them have to broken in nature; otherwise all quarks and leptons are 
massless, or the electroweak symmetry is not broken.  They have to 
broken by a very small amount.  This is where the dynamical symmetry 
breaking comes in.

\begin{figure}
	\centerline{
	\psfig{file=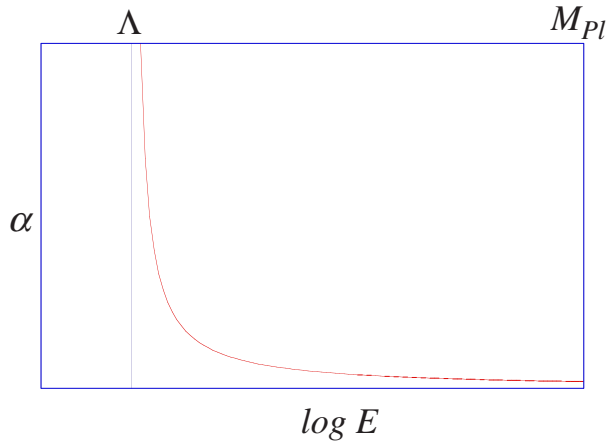,width=0.5\textwidth}
	}
	\caption{A schematic plot of the running coupling constant in an 
	asynmptotically gauge theory.}\label{fig:coupling}
\end{figure}

The idea of dynamical symmetry breaking is very simple.  Imagine an 
asymptotically free gauge theory, such as QCD. The QCD coupling constant is 
very small at high energies, {\it e.g.}\/, $\alpha_{s}(M_{\rm Planck}) 
\simeq 0.03$.  The coupling grows as the energy goes down, and 
eventually becomes infinite at the QCD scale $\Lambda \sim 200$~MeV. 
The QCD scale is a {\it derived}\/ energy scale from the initial 
coupling constant $\Lambda \sim M_{\rm Planck} 
e^{-2\pi/b_{0}\alpha_{s}(M_{\rm Planck})}$ where $b_{0}$ is the 
beta-function coefficient.  Because of the expontential factor of the 
perturbative coupling at the Planck scale, the QCD scale is {\it 
much}\/ smaller than the Planck scale.  When the QCD becomes strong, 
the chiral symmetry of the up, down, and strange quarks is broken 
dynamically, and a mass is generated for the proton.  This is why proton 
mass is so much smaller than the Planck scale; it truly {\it 
explains}\/ the hierarchy.  Put in more general terms, if a symmetry 
breaks due to an asymptotically free gauge theory becoming strong 
(Fig.~\ref{fig:coupling}), it is called dynamical symmetry breaking, 
and the scale of the symmetry breaking is naturally {\it much}\/ 
smaller than the fundamental energy scale.

Can we use this idea to explain the smallness of the supersymmetry 
breaking scale?  Dynamical breaking of supersymmetry, dynamicaly 
supersymmetry breaking, was proposed as a promising way to explain the 
hierarchy most clearly by Witten.\cite{Witten} To use this idea in 
realistic model building, however, we needed to understand 
non-perturbative dynamics of supersymmetric gauge theories.  Even 
though important progress had been made in mid-80's, the explosive 
progress had to wait until a series of works by Seiberg in mid-90's 
(see an excellent review by Intriligator and Seiberg\cite{Seiberg}).  
Now we have many uncontroversial models of dynamical supersymmetry 
breaking.  Probably the simplest model of dynamical supersymmetry 
breaking is a supersymmetric $SO(10)$ gauge theory with only one 
matter multiplet in the spinor representation {\bf 16}.\cite{SO10}

Now that we know that supersymmetry can be dymamically broken, the 
next question is how the particles in the supersymmetric standard 
model learn that supersymmetry is broken.  This is the issue of 
so-called ``mediation'' mechanism.  Once they acquire supersymmetry 
breaking effects, such as mass differences between the standard model 
particles and their superpartners at a few hundred GeV scale, 
$m_{H}^{2}$ can acquire a negative value and the electroweak symmetry 
can get broken.  On the other hand, phenomenology depends on the 
details of the mass spectrum, and hence on the details of the 
mediation mechanism.  In the past few years, there has been a tremendous 
progress in discovering new mediation mechanisms.

The simplest idea for mediation is to {\it do nothing}\/.  Even if 
there is apparently no interaction between the supersymmetric standard 
model and the sector which breaks supersymmetry dynamicaly (and is hence 
called ``hiden sector''), there is at least gravity.  If the dynamical 
supersymmetry breaking occurs at an energy scale $\Lambda_{SUSY}$, the 
gravitational effects induce supersymmetry breaking masses of scalar 
quarks and leptons at order $\Lambda^{2}_{SUSY}/M_{\rm Planck}$.  This 
option, however, had been regarded problematic, because the 
gravitational effects were believed to generate masses for gauginos 
(superpartners of gauge bosons) less than meV.\cite{Dine}  Recently, 
it was discovered that there is a quantum contribution to the gaugino 
masses due to the superconformal anomaly (``anomaly mediation'')\cite{anomaly}
\begin{equation}
	m_{\lambda} = \frac{\alpha}{2\pi} b_{0} 
	\frac{\Lambda_{SUSY}^{2}}{M_{\rm Planck}}.
\end{equation}
This is an interesting formula because it gives an unconventional mass 
spectrum among gauginos $M_{2} < M_{1} < M_{3}$, which leads to new 
phenomenology.

Another approach is to use the standard model gauge interactions 
themselves to mediate the supersymmetry breaking effects (``gauge 
mediation'').\cite{gauge} In this scheme, the sector which breaks 
supersymmetry dynamically at $\Lambda_{SUSY} \sim 10^{7}$~GeV is 
coupled to another sector called ``messenger sector'' by a new gauge 
interaction, which causes particles in the messenger sector to acquire 
masses, both supersymmetric and supersymmetry breaking ones at around 
$10^{5}$~GeV. The messenger particles carry the standard model gauge 
quantum numbers and their loops induce gaugino masses and scalar 
quark, lepton masses in the supersymmetric standard model at 
$10^{2}$--$10^{3}$~GeV. This is a beautiful mechanism that generates 
the scalar masses according to their gauge quantum numbers; it makes, 
for instance, $\tilde{d}$, $\tilde{s}$, $\tilde{b}$ degenerate and 
avoids the would-be flavor-changing effects by the superGIM mechanism.  
But aesthetically, having three separate somewhat decoupled sectors 
was unpleasant.  It turned out that the sector which breaks 
supersymmetry dynamically and the supersymmetric standard model can be 
coupled directly by the standard model gauge interactions (``direct 
gauge medation'').\cite{direct} Since these models predict definite 
superparticle spectra, they can be tested once superparticles are 
found.\footnote{After this talk, there appeared many more proposals on 
supersymmetry breaking.  See my brief review\cite{ICTP} for other 
ideas.}

\begin{figure}
	\centerline{
	\psfig{file=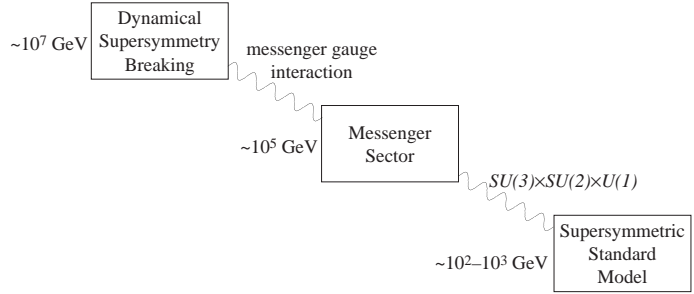,scale=0.3}
	}
	\caption{The original model of gauge mediation.\protect\cite{gauge}}
\end{figure}

\begin{figure}
	\centerline{
	\psfig{file=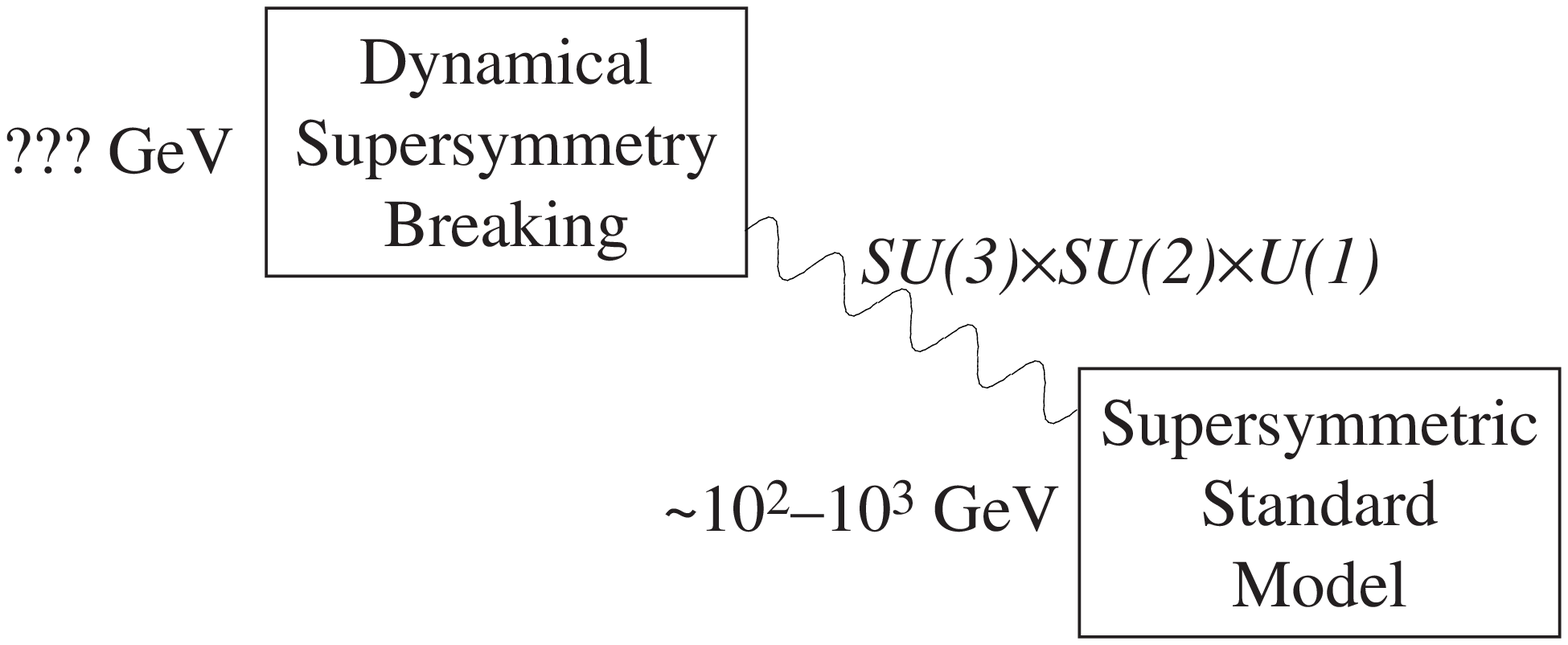,scale=0.3}
	}
	\caption{Direct gauge mediation models.\protect\cite{direct}}
\end{figure}

What about fermion mass hierarchy?  As discussed with the electron 
mass, smallness of the fermion masses is protected against self-energy 
corrections thanks to chiral symmetries.  This point can be used in 
model building.  The idea is to assume that some of the chiral 
symmetries (flavor symmetry) are indeed good, and hence first- and 
second-generation masses (possibly including $\tau$, $b$) are 
forbidden in the symmetry limit, while the top quark mass is allowed.  
Once the flavor symmetry is broken by a small amount, the lighter 
masses are generated.  A simple example is to have an approximate 
$U(1)$ symmetry, and assign the charges as given in 
Table~\ref{tab:anarchy} to the standard model particles.\cite{FN}  We 
assume that the flavor symmetry is broken by a small parameter 
$\epsilon(-1) \sim 0.04$ which carries the $U(1)$ charge $-1$.  Then 
one can write down the Yukawa matrices in power series expansion in 
$\epsilon$ so as to conserve the $U(1)$ charge,
\begin{equation}
	(Q_{1}, Q_{2}, Q_{3})
	\left( \begin{array}{ccc}
		\epsilon^{4} & \epsilon^{3} & \epsilon^{2} \\
		\epsilon^{3} & \epsilon^{2} & \epsilon \\
		\epsilon^{2} & \epsilon & 1 
	\end{array}
	\right)
	\left( \begin{array}{c}
		u_{1} \\ u_{2} \\ u_{3}
	\end{array}
	\right)
\end{equation}
for the up quarks, and similarly for the others.  Note that the 
symmetry requirements do not fix the precise coefficients in the power 
series in $\epsilon$, and there are unknown $O(1)$ constants in every 
matrix elements which we cannot predict.  But given that uncertainty, 
this pattern reproduces the observed quark, lepton masses and mixings 
including the solar and atmospheric neutrino oscillation data.  The 
idea can also be extended to non-abelian flavor symmetries which have 
virtues of suppressing supersymmetric flavor-changing effects by 
keeping (at least) first- and second-generation scalars degenerate 
(see, {\it e.g.}\/, my review on models\cite{YKIS} and references 
therein).

\begin{table}[t]
	\caption{A simple $U(1)$ flavor charge assignment which explains the 
	fermion masses and mixings.\protect\cite{anarchy}  The top row shows 
	the $SU(5)$ grand-unified multiplets.\label{tab:anarchy}}
	\centerline{
	\begin{tabular}{|c|ccc|cc|c|}
		\hline
		& \multicolumn{3}{c|}{{\bf 10}} & \multicolumn{2}{c|}{\bf 5}
		& {\bf 1}\\
		generation & $Q$ & $u^{c}$ & $e^{c}$ & $L$ & $d^{c}$ & $N^{c}$\\
		\hline
		1st & \multicolumn{3}{c|}{$+2$} & \multicolumn{2}{c|}{$0$} & $0$\\
		2nd & \multicolumn{3}{c|}{$+1$} & \multicolumn{2}{c|}{$0$} & $0$\\
		3rd & \multicolumn{3}{c|}{$0$} & \multicolumn{2}{c|}{$0$} & $0$\\
		\hline
	\end{tabular}
	}
\end{table}

The approximate flavor symmetry hence allows a successful model 
building, but it does not explain the origin of the small flavor 
symmetry breaking parameter ($\epsilon \sim 0.04 \ll 1$ in the above 
example).  There are various ideas to explain the origin of the small 
number in this context.  One of them is to use radiative breaking 
$\epsilon \sim g^{2}/16\pi^{2}$.  This one-loop factor arises in the 
context of anomalous $U(1)$ gauge symmetry in string theory\cite{FY} 
or ordinary loop factor in the gauge mediated models.\cite{heresy}
The other possibility is to generate the small number dynamically as 
in the case of supersymmetry breaking.\cite{CHM}  It may also be due 
to the compositeness of {\bf 10} multiplets.\cite{composite}

It is noteworthy that the supersymmetric models with approximate 
flavor symmetry allow interesting flavor-changing effects.  For 
instance, among two generations of down quarks, the mass matrix is 
given by
\begin{equation}
	(Q_{1}, Q_{2})
	\left( \begin{array}{cc}
		m_{d} & m_{s} \lambda \\
		m_{s} \lambda & m_{s}
	\end{array}
	\right)
	\left( \begin{array}{c}
		d_{1} \\ d_{2}
	\end{array}
	\right),
\end{equation}
where the hierarchy $m_{d} \ll m_{s} \lambda \ll m_{s}$ with 
$\lambda \sim 0.22$ may be a consequence of a $U(1)$ flavor symmetry.
The corresponding mass matrix for the squarks then would be
\begin{equation}
	m_{SUSY} (\tilde{Q}_{1}, \tilde{Q}_{2})
	\left( \begin{array}{cc}
		a m_{d} & b m_{s} \lambda \\
		c m_{s} \lambda & d m_{s}
	\end{array}
	\right)
	\left( \begin{array}{c}
		\tilde{d}_{1} \\ \tilde{d}_{2}
	\end{array}
	\right),
\end{equation}
where the flavor symmetry does not fix the $O(1)$ coefficients $a$, 
$b$, $c$, $d$.  Going to the basis where the quark mass matrix is 
diagonalized by the Cabibbo rotation, the squark mass matrix is also 
rotated by the same amount and we find the new mass matrix in this 
basis
\begin{equation}
	m_{SUSY} (\tilde{Q}_{1}, \tilde{Q}_{2})
	\left( \begin{array}{cc}
		(a-b-c+d) m_{d} & (b-d) m_{s} \lambda \\
		(c-d) m_{s} \lambda & d m_{s}
	\end{array}
	\right)
	\left( \begin{array}{c}
		\tilde{d}_{1} \\ \tilde{d}_{2}
	\end{array}
	\right).
\end{equation}
Unless there is any particular reason for $b=c=d$, there remains 
off-diagonal elements in the squark mass matrix, which feeds into 
flavor-changing loop diagrams (see Gabbiani {\it et al}\/\cite{Gabbiani} 
for an extensive analysis of such flavor-changing effects).  For 
instance, it may contribute to $\epsilon'/\epsilon$ at an interesting 
level,\cite{MM}
\begin{equation}
	\frac{\epsilon'}{\epsilon} \sim 3 \times 10^{-3}
		\left( \frac{500~{\rm GeV}}{m_{SUSY}}\right)^{2}
		\Im (b-a).
\end{equation}
Similar contributions appear in neutron and electron electric dipole 
moments and $\mu \rightarrow e\gamma$.

One interesting question is how exactly does a symmetry break 
dynamically in gauge theories.  For instance in QCD, it is known that 
the $SU(3)_{L} \times SU(3)_{R}$ flavor symmetry is broken dynamically 
to $SU(3)_{V}$, and the corresponding Nambu--Goldstone bosons are the 
octet of pseudo-scalar mesons $\pi^{\pm}$, $\pi^{0}$, $K^{\pm}$, 
$K^{0}$, $\overline{K}^{0}$, $\eta$.  Before discussing microscopic 
mechanism for the flavor symmetry breaking, let me remind you of a 
well-known qualitative argument for confinement in gauge theories due 
to monopole condensation by `t Hooft and Mandelstam.\cite{monopole}  
In Type-II superconductors, there is a condensate of Cooper pairs 
(electric condensate), which causes magnetic fields to be squeezed in 
flux tubes.  If you could place a pair of magnetic monopoles inside a 
superconductor, the magnetic field flux should be in a flux tube 
stretched between the monopoles and hence the energy increases 
linearly with the distance.  This is nothing but the confinement.  If 
you interchange electric and magnetic everywhere, you find that, in a 
system with a magnetic monopole condensate, the electric fields 
between charged particles are squeezed in flux tubes and the potential 
energy increases linearly with distance, and hence the electric 
charges are confined.  This mechanism was shown to be operative at 
least in $N=2$ supersymmetric theories in a beautiful analysis by 
Seiberg and Witten.\cite{SW}  When there are quarks coupled to these 
theories, it turns out that the magnetic monopoles acquire flavor 
quantum numbers.  For instance in $Sp(n_{c})$ gauge theories with 
$n_{f}$ quarks, the magnetic monopoles belong to the spinor 
representation of the $SO(2n_{f})$ flavor symmetry.  When they 
condense, not only they cause confinement, but also break the flavor 
symmetry from $SO(2n_{f})$ to $U(n_{f})$.\cite{CKM}

\section{Conclusions}

I have hopefully convinced you that the dynamical symmetry breaking is 
a very natural concept in gauge theories and is useful in explaining 
hierarchies.  Since the dynamics of supersymmetric gauge theories is 
understood quite well by now, we can use them to construct realistic, 
asthetically appealing models.  When applied to mechanims of 
supersymmetry breaking and flavor symmetry breaking, different models 
predict different superparticle spectra and flavor-changing effects, 
and hence can be tested experimentally.

\section*{Acknowledgments}
I thank the patience of organizers waiting for my writeup.  This work 
was supported in part by the Department of Energy under contract 
DE--AC03--76SF00098, and in part by the National Science Foundation 
under grant PHY-95-14797.

\end{document}